# Evidence-based Explanation to Promote Fairness in AI systems

**Juliana Jansen Ferreira**
IBM Research
Rio de Janeiro, RJ, Brazil
jjansen@br.ibm.com

**Mateus de Souza Monteiro**
IBM Research
Rio de Janeiro, RJ, Brazil
msmonteiro@ibm.com

## Abstract

As Artificial Intelligence (AI) technology gets more intertwined with every system, people are using AI to make decisions on their everyday activities. In simple contexts, such as Netflix recommendations, or in more complex context like in judicial scenarios, AI is part of people's decisions. People make decisions and usually they need to explain their decision to others or in some matter. It is particularly critical in contexts where human expertise is central for decision-making. In order to explain their decisions with AI support, people need to understand how AI is part of that decision. When considering the aspect of fairness, the role that AI has on a decision-making process becomes even more sensitive since it affects the fairness and the responsibility of those people making the ultimate decision. We have been exploring an evidence-based explanation design approach to 'tell the story of a decision'. In this position paper, we discuss our approach for AI systems using fairness sensitive cases in the literature.

**CSS Concepts**
• **Human-centered computing~Human computer interaction (HCI)**;

## Introduction
The use of Artificial Intelligence (AI) on practically every technology-based task people perform is a reality. From making simple decisions, such as what movie to watch[1], to complex ones like a judge ruling on someone's freedom [1][4], AI is already an influential player on the decision-making process. Even without direct action from a person, an algorithmic decision-making can affect that person's life [2], and that person may not even know it.

In decision-making processes where people must have the final word, the role AI plays on that decisions needs to be very clear for decision-makers. AI is used as an empowerment tool for people [2], but if an AI output can be part of the decision, decision-makers need to understand how AI got to that output. In this scenario, the demand for AI explanation just rises and becomes imperative for decisions people make that might have impact on other people's lives [2][13].

Explainable AI is a hot topic due to the AI reach on everyday technologies and have been investigated by different perspectives [6][13][14]. When the decision-making process relays on human expertise, the use of pieces of evidence to explain the decisions is a way to show experts' rational to reach the decision. An evidence is any data or item of information that is relevant regarding some matter to be proven or disproven, it is data that support statements [5].

---
[1] https://research.netflix.com/research-area/machine-learning

As AI explanation, fairness in AI is not a one-dimension characteristic. For now, we consider two dimensions: context and stakeholders of AI fairness. First, fairness is context-dependent, and it needs to deal with more scenarios than just nondiscrimination, as covered by statistical fairness [8]. Second, fairness needs to account for all people involved with the AI system, directly or not. The notion of fairness varies for different stakeholders [9] and do not always correspond to mathematical interpretations of algorithmic fairness or biases [11][12].

We have been investigating the context of a knowledge-intensive decision-making process where decisions are based on human expertise [7]. In this context, we have an AI system to support de decision process and we have been investigating an evidence-based explanation to support decision-makers on telling the story of their decision. We believe that our approach to decision rational explanation with pieces of evidence can be used to promote fairness in the decision process, especially to highlight where the AI system provides input for that decision. In this position paper, we discuss our investigation on evidence-based explanation for AI systems, considering fairness in literature cases.

## Fairness and AI
Law is a particular rich context to explore the topic of AI fairness. Notwithstanding the potential benefits of AI for law, given the concerns of affecting people's liberty, safety, or privacy, opaque algorithms have come under criticism [1]. The failure to present what is behind the results hollow people's sense of fairness and trust. Reasons for eXplainable AI (XAI) in courts are abundant, such as ([1], pp. 1845-1846).

Table 1. Questions for XAI in decision-making [4]

| Questions for XAI in decision-making: |
|---|
| **Q1:** What are the main factors in a decision? *A list of pieces of evidence that went into a decision;* |
| **Q2:** Would be changing a certain factor have changed the decision? *By looking at the effect of changing that information on the output and comparing it to our expectations, we can infer whether it was used correctly;* |
| **Q3:** Why did two similar-looking cases get different decisions? *Whether a specific factor was determinative to another decision. Useful to assess the integrity of the decision-maker.* |

Veale, Kleek, and Binns [10] argue that improvement in the fairness challenges for criminal justice and courts is more than just algorithms transparency. It will only happen through close collaboration with different disciplines, practitioners, and affected stakeholders.

Deek [1] suggests that courts should focus on two principles: maximizing xAI's ability to help identify errors and biases within the algorithm, and aligning the form of xAI in a given case with the needs of the relevant audiences. Besides, Doshi-Velez and Mason ([4], pp. 2-3) affirm that the explanation should be able to answer a set of questions (Table 1).

Fairness and explanation are strongly dependent. The explanation of the decision process is a way to guarantee fairness to all people impacted by AI-related decisions. Although the complexity of defining fairness, Rudin, Wager, and Coker [3] summarizes by arguing that it is not fair life-changing decisions made by unclear, untrusted, and unverifiable explanation.

## Discussion

Our evidence-based explanation design approach aims to support experts to tell their decision story. Every step of the decision is presented is a timeline with the related evidence, to allow the decision-maker to revisit his rational considering each portion of that final decision (Figure 1). This story-based design approach can be used to assess fairness in the decision process with AI.

The events in the timeline might present the AI systems as a player that needs feedback interaction with the expert (Figure 1-B), but also indicate events where expert's tacit knowledge was the main evidence in different ways: a) as new input (Figure 1-A), by associating previous data as new knowledge (Figure 1-C), and associating tacit knowledge to previous data (Figure 1-D). The decision itself is a event that composes the story.

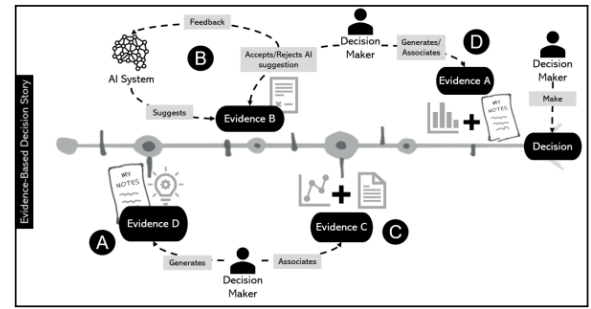

Figure 1. Decision-Making Story based on evidence

Our approach aligns with the set of question for explanation proposed by Doshi-Velez and Mason [4] (Table 1). It allows the decision-maker or other stakeholders to check the factors for the decision, to inquire if different factors could impact the decision, and compare different decisions' stories. The decision-maker himself is part of the story, is a relevant factor of the decision.

Considering the context and the stakeholders of the explanation might promote fairness in the decision process supported by AI. The receiver of the explanation and the reason for that explanation will define what kind of evidence can aid fairness for all people involved in the decision story. Can an algorithmic explanation provide fairness to stakeholders? Sometimes yes, but, sometimes no.

## References

[1] Ashley Deeks. 2019. The Judicial Demand for Explainable Artificial Intelligence. Columbia Law Review 119, 7: 1829–1850.

[2] Ashraf Abdul, Jo Vermeulen, Danding Wang, Brian Y. Lim, and Mohan Kankanhalli. 2018. Trends and Trajectories for Explainable, Accountable and Intelligible Systems: An HCI Research Agenda. Proceedings of the 2018 CHI Conference on Human Factors in Computing Systems, ACM, 582:1–582:18.

[3] Cynthia Rudin, Caroline Wang, and Beau Coker. 2018. The age of secrecy and unfairness in recidivism prediction. arXiv preprint arXiv:1811.00731.

[4] Finale Doshi-Velez, Mason Kortz, Ryan Budish, et al. 2017. Accountability of AI under the law: The role of explanation. arXiv preprint arXiv:1711.01134.

[5] Gheorghe Tecuci, David A. Schum, Dorin Marcu, and Mihai Boicu. 2016. Intelligence analysis as discovery of evidence, hypotheses, and arguments: connecting the dots. Cambridge University, New York NY.

[6] Jonathan Dodge, Q. Vera Liao, Yunfeng Zhang, Rachel KE Bellamy, and Casey Dugan. 2019. Explaining Models: An Empirical Study of How Explanations Impact Fairness Judgment. arXiv preprint arXiv:1901.07694.

[7] Juliana Jansen Ferreira, Ana Fucs, and Vinícius Segura. 2019. Should I Interfere' AI-Assistants' Interaction with Knowledge Workers: A Case Study in the Oil and Gas Industry. Extended Abstracts of the 2019 CHI Conference on Human Factors in Computing Systems, ACM, CS20:1–CS20:7.

[8] Kenneth Holstein, Jennifer Wortman Vaughan, Hal Daumé III, Miro Dudik, and Hanna Wallach. 2019. Improving fairness in machine learning systems: What do industry practitioners need? Proceedings of the 2019 CHI Conference on Human Factors in Computing Systems, 1–16.

[9] Kenneth Holstein, Jennifer Wortman Vaughan, Hal Daumé III, Miro Dudik, and Hanna Wallach. 2019. Improving fairness in machine learning systems: What do industry practitioners need? *Proceedings of the 2019 CHI Conference on Human Factors in Computing Systems*, 1–16.

[10] Michael Veale, Max Van Kleek, and Reuben Binns. 2018. Fairness and accountability design needs for algorithmic support in high-stakes public sector decision-making. Proceedings of the 2018 chi conference on human factors in computing systems, 1–14.

[11] Min Kyung Lee and Su Baykal. 2017. Algorithmic mediation in group decisions: Fairness perceptions of algorithmically mediated vs. discussion-based social division. Proceedings of the 2017 ACM Conference on Computer Supported Cooperative Work and Social Computing, 1035–1048.

[12] Rachel KE Bellamy, Kuntal Dey, Michael Hind, et al. 2018. AI Fairness 360: An extensible toolkit for detecting, understanding, and mitigating unwanted algorithmic bias. arXiv preprint arXiv:1810.01943.

[13] Tim Miller. 2019. Explanation in artificial intelligence: Insights from the social sciences. Artificial Intelligence 267: 1–38.

[14] Yogesh K. Dwivedi, Laurie Hughes, Elvira Ismagilova, et al. 2019. Artificial Intelligence (AI): Multidisciplinary perspectives on emerging challenges, opportunities, and agenda for research, practice and policy. International Journal of Information Management: S026840121930917X.